\documentclass[fleqn,usenatbib,useAMS]{mnras}

\usepackage{graphicx}	
\usepackage{amsmath}	
\usepackage{amssymb}	
\usepackage{multicol}        
\usepackage{bm}		
\usepackage{pdflscape}	
\usepackage{natbib, color, bm, epsfig}
\usepackage{widetext}
\usepackage{array}
\usepackage{makecell}
\usepackage{xcolor}

\usepackage[T1]{fontenc}
\usepackage{ae,aecompl}
\usepackage{newtxtext,newtxmath}

\newcommand{\Mach}{M_{\text{3D}}}

\newcommand{\D}{\mathrm{d}}
\newcommand{\fsh}{f_{\text{sh.}}}
\newcommand{\Max}{\mathcal{M}_t}

\title[Density-Velocity Correlations]{Density and Velocity Correlations in Isothermal Supersonic Turbulence}

\author[B. Rabatin et al.]{Branislav Rabatin,$^{1}$%
\thanks{Contact e-mail: \href{mailto:br18b@fsu.edu}{br18b@fsu.edu}}%
David C. Collins $^{1}$
\\
$^{1}$Florida State University, Tallahassee, FL 32309}

\begin{document}
\label{firstpage}
\pagerange{\pageref{firstpage}--\pageref{lastpage}}
\maketitle

\begin{abstract}
In star-forming clouds, high velocity flow gives rise to large fluctuations of density.  In this work we explore the correlation between velocity magnitude (speed) and density.  We develop an analytic formula for the joint probability distribution (PDF) of density and speed, and discuss its properties.  In order to develop an accurate model for the joint PDF, we first develop improved models of the marginalized distributions of density and speed.  
We confront our results with a suite of 12 supersonic isothermal simulations with resolution of $1024^3$ cells in which the turbulence is driven by 3 different forcing modes (solenoidal, mixed and compressive) and 4 r.m.s. Mach numbers (1, 2, 4, 8).
We show, that for transsonic turbulence, density and speed are correlated to a considerable degree and the simple assumption of independence fails to accurately describe their statistics. In the supersonic regime, the correlations tend to weaken with growing Mach number.
Our new model of the joint and marginalized PDFs are a factor of 3 better than uncorrelated, and provides insight into this important process.
\end{abstract}




\begin{figure*} \begin{center}
\includegraphics[width=0.99\textwidth]{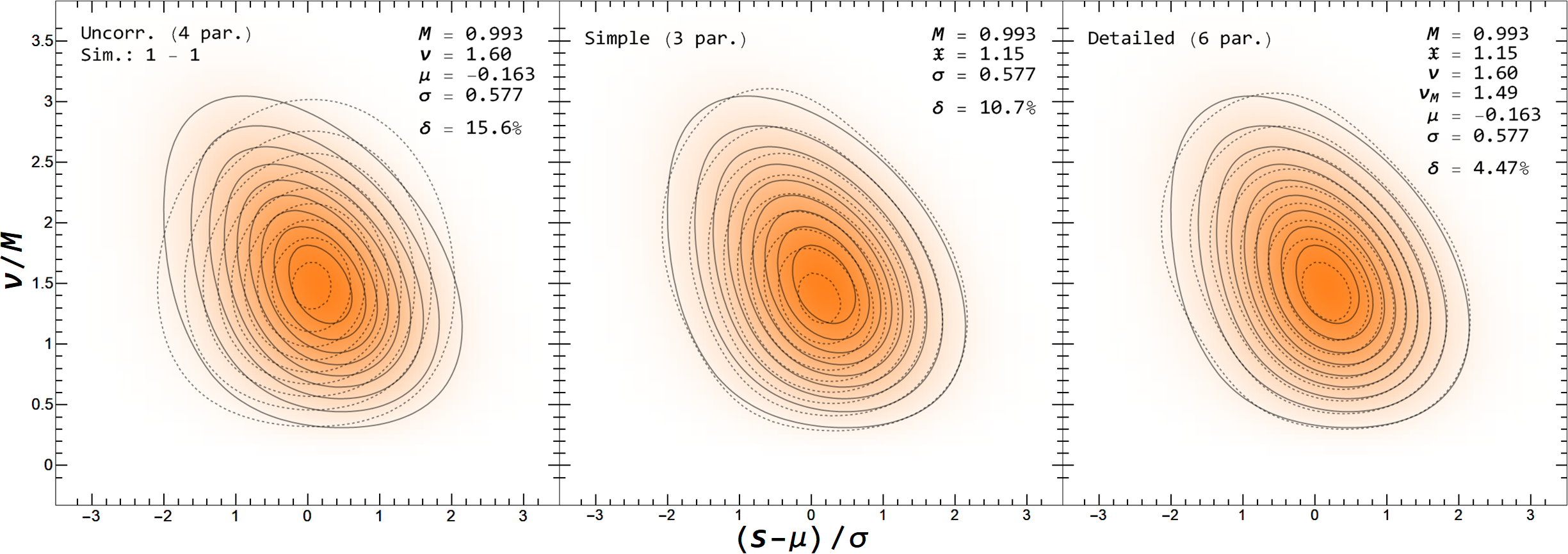}
\caption[ ]{The joint PDF of speed $v$ vs. log density $s$ (color, solid lines) along with models of the PDF (dashed lines).  The left panel includes the joint PDF assuming uncorrelated density and speed. The middle panel shows our simple model that includes correlations, and the right panel shows our detailed model for the PDF that includes correlations and improved models for the marginalized PDFs of density and speed.}
\label{fig:teaser} \end{center} \end{figure*}

\section{Introduction}

Star-forming clouds of molecular hydrogen, which are known to be undergoing turbulent supersonic motion, are often modeled as isothermal in astrophysical simulations. This approximation is facilitated by rapid cooling rates of the molecular clouds \citep{Armstrong95, Elmegreen04, BigProblems, PadoanSFR14}, which keeps the temperature roughly constant.

This reasonably simple yet powerful model is capable of explaining the observed density fluctuations within the molecular clouds, which can be used to predict many properties of star formation, such as the star formation rate \citep{Krumholz05, Padoan11, Hennebelle11, Federrath12} and the initial stellar mass distribution \citep{Padoan02}. While supersonic turbulent motion inhibits the collapse and star formation by increasing the effective Jeans mass, at the same time it gives rise to large density variations allowing for a local collapse \citep{MacLow04}.

The interplay between density and velocity fluctuations is fundamental to understanding star formation \citep{Federrath10}. Describing the statistics of the fundamental dynamical quantities including the correlations between them reveals the statistical behavior of all derived quantities, including kinetic energy and the joint PDF of kinetic and thermal energy.

The main purpose of this work is to explore $f_{sv}(s,v)$, the joint probability distribution
function (PDF) between the log of density, $s=\log \rho$, and speed, $v$. The simplest
assumption is that $s$ and $v$ are independent of one another, in which case the
joint distribution is the product of the marginalized distributions:
\begin{align}
    f_{(s,v)}&= f_s(s) f_v(v) \\
    f_s(s)&=\int \limits_{-\infty}^\infty \D v \, f_{(s,v)}(s,v)\\
    f_v(v)&=\int \limits_0^\infty \D s \, f_{(s,v)}(s,v).
\end{align}
The density PDF is typically treated as lognormal, $f_s(s) = \mathcal{N}(s;\mu, \sigma)$, a Gaussian $\mathcal{N}$ with mean $\mu$ and variance $\sigma$.
Speed, $v$, is usually modeled with a Maxwellian distribution; $f_v (v) = \mathcal{M} (v; M)$ with the 1D Mach number $M = \sqrt{\langle v^2 \rangle / 3}$. In this work, we improve on all three assumptions.
The finite shock model \citep{fsh23} as an extension of a simple Gaussian PDF of density is discussed in Section \ref{section:density}. In Section \ref{section:speed} we introduce a tilted Maxwellian to better fit the statistics of speed.  Finally, we find a correction to the joint PDF in Section \ref{section:joint}.

Figure \ref{fig:teaser} shows three models for the joint distribution along with simulated data. The color and solid contours are taken from simulations
described in Section \ref{section:methods}. In the left panel, the dashed contours show the simple assumption of uncorrelated variables. Clearly the
shape of the model does not agree with the simulated data. The second panel shows our first correction to the joint PDF, which introduces a correlation between density and speed, but continues to assume a lognormal for density and Maxwellian for speed. The third panel shows our detailed model, with the corrected joint PDF and improved density and speed PDFs.

An important aspect of this work is the lack of fitting of any kind. All of the results come from moments of the data, and not by fitting a model to the simulated histograms.

The paper is organized as follows. In Section \ref{section:methods} we discuss the code, simulations, and analysis. In
Section \ref{section:density} we describe the finite shock model for the density
PDF. In Section \ref{section:speed} we discuss our updated distribution of speed.  In Sections \ref{section:joint} and \ref{section:specific} we show our new joint
distribution.  In Section \ref{section:moments} we show that our model works
well even for higher order moments of the distribution.  Finally we conclude in
Section \ref{section:conclusions}.

\section{Methods}
\label{section:methods}

The suite of numerical simulations was performed using the hydrodynamic code Enzo \citep{Bryan14} using the piecewise parabolic method \citep{Woodward84}. The simulation domain consists of a cube of unit length with periodic boundary conditions. Each simulation is described by two parameters, the forcing mode $\xi$ and Mach number $M$, both introduced via the Stochastic forcing module implemented within Enzo (Schmidt, Federrath, 2008). The forcing mode $\xi \in [0,1]$ is the weight of the solenoidal components of the forcing field. The value of $\xi = 0$ corresponds to the purely compressive forcing field, whereas $\xi = 1$ represents the purely solenoidal forcing. 
The target mach number is achieved by adding energy at the large scale at a rate equal to the Mach-number dissipation rate, $\epsilon M^3/L$ \citep{MacLow04}.


For each Mach number $M$ we consider the turnover scale $\tau$ as the time scale at which two frames become statistically uncorrelated. The turnover time is roughly equal to the turbulent crossing time $\tau_{\text{turb.}} = (L/2)/M$, where $L$ is the size of the box with $L/2$ being the size of the driving pattern and $M$ is the 1D r.m.s. Mach number, $M = \sqrt{\langle v^2 / 3 \rangle}$. Each simulation is run for $9 \tau$ with the step of $0.1 \tau$. For statistical purposes, only frames with $t \geq 2 \tau$ are considered, as the fluid settles in its chaotic turbulent motion. Thus 71 snapshots of statistics within each simulation. This approach to obtain statistical data is common in similar astrophysical simulations \citep{Porter99, Porter00, Federrath10, Federrath13, Federrath21}.

The simulation grid consists of $N = 1024^3$ cells with each cell $\ell$ containing the same volume $\delta V_\ell = 1/1024^3$.  Our suite of simulations employed 1D r.m.s. Mach numbers 1, 2, 4, 8, and three values of the forcing parameter, $\xi = 0, 1/2, 1$.

Table \ref{tab:simpars} describes the simulations and the resulting parameters.  The first column names the simulation by way of forcing parameter and target Mach number.  The second column shows the actual 1d Mach number realized by the simulation.  The third column shows the ratio of volume-weighted Mach number to mass-weighted Mach number, $\mathfrak X$. The following two columns show the volume-weighted mean speed $\langle v \rangle$ and its mass-weighted counterpart $\langle \rho v \rangle$. The final three columns show the volume-weighted mean and variance of $s$, $\mu$ and $\sigma$, and the number of shocks. 

\begin{center}
    \begin{table*}
\begin{tabular}{| c || c | c | c | c || c | c | c |}
\hline
Sim. ($\xi-M$) & $M$ & $\mathfrak X$ & $\left\langle v \right\rangle$ & $\left\langle \rho v \right\rangle$ & $\mu$ & $\sigma$ & $n$ \\ \hline \hline
$0-1$ & $0.977$ & $1.35$ & $1.55$ & $1.33$ & $-1.07$ & $1.59$ & $6.45$ \\
$0-2$ & $1.99$ & $1.38$ & $3.21$ & $2.72$ & $-2.32$ & $2.48$ & $4.43$ \\
$0-4$ & $3.92$ & $1.43$ & $6.34$ & $5.25$ & $-3.61$ & $3.29$ & $2.84$ \\
$0-8$ & $7.77$ & $1.41$ & $12.6$ & $10.5$ & $-4.73$ & $3.93$ & $2.35$ \\ \hline
$1/2-1$ & $0.999$ & $1.14$ & $1.61$ & $1.50$ & $-0.196$ & $0.634$ & $53.6$ \\
$1/2-2$ & $2.00$ & $1.21$ & $3.23$ & $2.92$ & $-0.618$ & $1.14$ & $50.8$ \\
$1/2-4$ & $3.98$ & $1.16$ & $6.41$ & $5.91$ & $-1.15$ & $1.60$ & $16.0$ \\
$1/2-8$ & $7.89$ & $1.14$ & $12.7$ & $11.8$ & $-1.57$ & $1.91$ & $11.0$ \\ \hline
$1-1$ & $0.993$ & $1.15$ & $1.60$ & $1.49$ & $-0.163$ & $0.577$ & $70.3$ \\
$1-2$ & $1.98$ & $1.16$ & $3.20$ & $2.96$ & $-0.497$ & $1.01$ & $291$ \\
$1-4$ & $3.71$ & $1.14$ & $5.99$ & $5.58$ & $-0.883$ & $1.37$ & $53.8$ \\
$1-8$ & $8.04$ & $1.13$ & $13.0$ & $12.2$ & $-1.24$ & $1.68$ & $13.0$ \\
\hline
\end{tabular}
\caption[] {Simulation parameters. The first column denotes each simulation in the form of $\xi-M$ where $\xi$ is the forcing mode and $M$ is the nominal 1D r.m.s. Mach number. The second column lists the measured 1D r.m.s. Mach number. The third column represents $\mathfrak X = \langle v^2 \rangle / \langle \rho v^2 \rangle$, the ratio between the volume- and mass-weighted Mach numbers, squared. Columns 4 and 5 show the volume- and mass-weighted mean values of speed, respectively. Columns 6, 7, 8 show the statistical parameters of density; mean $\mu = \langle s \rangle$, standard deviation $\sigma = \sqrt{\langle s^2 \rangle - \langle s \rangle^2}$ and the number of shocks, $n$, using equation \eqref{eq:nV_from_mu_sigma}.}
\label{tab:simpars}
\end{table*}
\end{center}

\subsection{Analysis}\label{sec:weighting}

The probability distribution function, $f_Q(q)$, for a random quantity, $Q$, is
the probability that $Q$ will realize a value within the interval $[q,q+dq]$.
This can be found as
\begin{align}
    f_Q(q) =\frac{1}{V} \int_V d^3 x \, \delta( q-Q(\vec{x})),
\end{align}
where $V$ is the volume of the sample.

We can alternatively weight our PDF with other quantities, $W$, as
\begin{align}
    f^{(W)}_Q(q) = \frac{1}{W_{\text{net}}} \int_V d^3 x \, W(\vec{x}) \, \delta( q -
Q(\vec{x})),
\end{align}
where $W_{\text{net}}$ is the total of $W$ on the domain. This is useful as it gives
an alternative view of the variable.

We will find it valuable to explore weighting by volume $(V)$, mass $(M)$, and kinetic
energy $(E)$. 2D PDFs weighted by different quantities are related to one another by the following useful formulae:
\begin{align}
    \label{eq:PDF_weighting_MV2D}
    f^{(M)}_{(s,v)} (s, v) &= e^s f^{(V)}_{(s,v)} (s, v) \\
    \label{eq:PDF_weighting_EV2D}
    f^{(E)}_{(s,v)} (s, v) &= \frac{e^s v^2}{\langle e^s v^2 \rangle} f^{(V)}_{(s,v)} (s, v) \\
    \label{eq:PDF_weighting_EM2D}
    f^{(E)}_{(s,v)} (s, v) &= \frac{v^2}{\langle e^s v^2 \rangle} f^{(M)}_{(s,v)} (s, v)
\end{align}

For 1D PDFs, the only simple analytic expressions possible are the following
\begin{align}
    \label{eq:PDF_weighting_MV1D}
    f^{(M)}_{s} (s) &= e^s f^{(V)}_{s} (s) \\
    \label{eq:PDF_weighting_EM1D}
    f^{(E)}_{v} (v) &= \frac{v^2}{\langle e^s v^2 \rangle} f^{(M)}_{v} (v).
\end{align}
Relationships between other weights and quantities, e.g., $f_v^{(M)}(v)$ and $f_v^{(V)}(v)$, are only possible by integrating the joint distributions.

The ratio of volume-weighted Mach number to its mass-weighted counterpart will prove to be a useful quantity:
\begin{equation}
\mathfrak X = \frac{\langle v^2 \rangle}{\langle e^s v^2 \rangle} \label{eq:defX} = \frac{M^2}{M_M^2}
\end{equation}
which serves as a loose measure of the correlation between density and velocity. Here we have introduced the mass-weighted Mach number, $M_M = \sqrt{\langle \rho v^2 \rangle/3}$.

For the purposes of numerically comparing histograms binned from data, $f^{\text{(data)}}$, with a theoretical model $f^{\text{(theory)}}$ we employ the $L_1$ norm
\begin{equation}
    \delta = \sum_{\text{bin $b$}} \left| f^{\text{(data)}}_b - f^{\text{(theory)}} (b_{\text{cen.}}) \right|  |b|
\end{equation}
where the model function is evaluated at the bin center $b_{\text{cen.}}$ and $|b|$ indicates the bin measure (length, area, volume, ...). This formula closely mimics the analogous integral $L_1$ norm.
\begin{figure*} \begin{center}
    \includegraphics[width=0.99\textwidth]{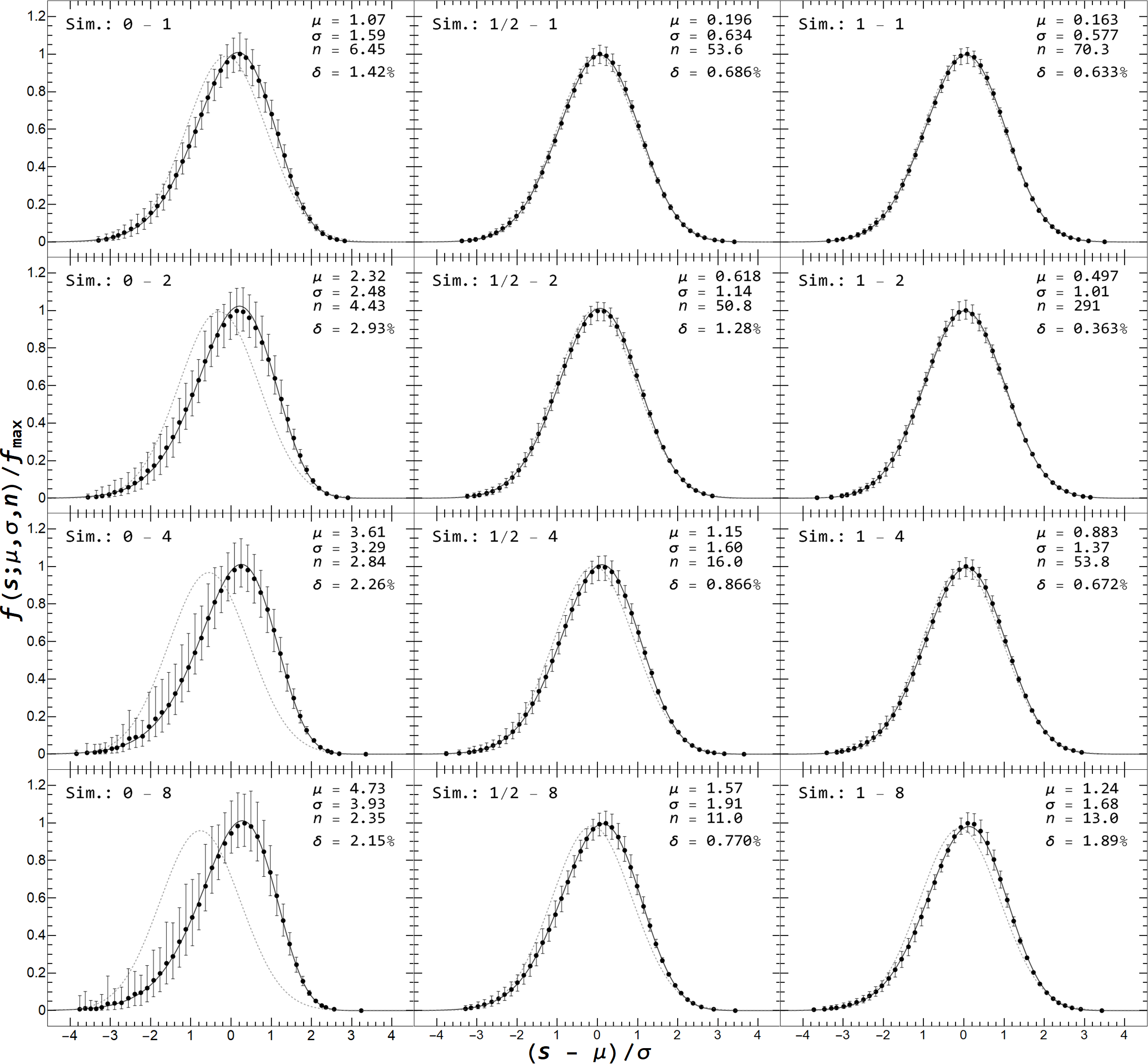}
\caption[ ]{Plots of the PDF (vertical axis) of log density (horizontal axis). For the sake of clarity, the horizontal axis is shifted towards the center and scaled by the width of the histogram, and the vertical axis is scaled by the maximum of the distribution. Data points (dots) along with the error bars (vertical lines) are represented in black. The black line represents the finite shock model function with parameters listed on the plot. Parameters $\mu$ and $\sigma$ are measured as ensemble averages, while $n$ is estimated using equation \eqref{eq:nV_from_mu_sigma}. Dashed gray lines depict the ideal Gaussian function using equation \eqref{eq:lognormal} whose only parameter is $\sigma$.}
\label{fig:logrhoVfits} \end{center} \end{figure*}
\section{Density in supersonic isothermal turbulence}\label{section:density}

The knowledge of the statistical properties of density within the star-forming clouds is one of the cornerstones of many star formation theories \citep{PNJ97, Krumholz05, Padoan11, Hennebelle11, Federrath12, BigProblems}. A turbulent medium without self-gravity can be shown to exhibit near lognormal density fluctuations, a result of the self-similar statistics of isothermal, supersonic flows \citep{Vazquez-Semadeni94, PNJ97, NordlunPadoan99, Passot98, Federrath08, Schmidt09}, later also extended to flows magnetized with ideal MHD \citep{Molina12}. In the scope of isothermal turbulence the PDF of log density $s = \log \rho / \rho_0$ can be approximated by a Gaussian
\begin{equation}
    f_s (s; \sigma) = \mathcal{N} (s; - \sigma^2 / 2, \sigma) = \frac{1}{\sqrt{2 \pi \sigma^2}} \exp \left( - \frac{\left( s + \sigma^2 / 2 \right)^2}{2 \sigma^2} \right)
    \label{eq:lognormal}
\end{equation}
with variance $\sigma^2 = \langle s^2 \rangle - \langle s \rangle^2$ and mean value $\mu = \langle s \rangle = - \sigma^2 / 2$ that fixes the mean density, $\langle e^s \rangle = 1$. In the longormal approximation, the variance is known to depend on the r.m.s. sonic Mach number $\Mach = \sqrt{\langle v^2 \rangle}$ and the weight of the solenoidal components of the forcing, $\xi$; $\sigma^2 \approx \log \left( 1 + b^2 \Mach^2 \right)$ \citep{Padoan11}.

While the lognormal approximation already provides a reasonably accurate picture of the density fluctuations, several works propose various corrections to the PDF of density, either purely within the context of turbulence \citep{Hopkins13, Squire17, Mocz19, fsh23}, or due to other phenomena extending beyond the framework of isothermal turbulence \citep{Scalo98, OGS99, Klessen_gravity20}.

In this work we make use of the finite shock model of density fluctuations \citep{fsh23}, that describes the PDF of log density $s$ arising from a series of shocks traversing the turbulent medium, each adjusting the local density by a factor proportional to the local sonic Mach number, drawn from an idealized Maxwell distribution. When the number of the shocks grows to infinity, the PDF of density approaches a lognormal. However, for a finite number of shocks $n$, the distribution in $s$ can be described via its characteristic function, $\phi$
\begin{equation}
    \begin{aligned}
	\fsh (s; \mu, \sigma, n) &= \frac{1}{\sigma} \int \limits_{- \infty}^\infty \D \omega \, \phi (\omega; n) \exp \left( - i \omega \frac{s - \mu}{\sigma} \right) \\
    \end{aligned}
    \label{eq:fshock_precise}
\end{equation}
where the parameters $\mu \equiv \left\langle s \right\rangle$ and $\sigma^2 \equiv \left\langle s^2 \right\rangle - \mu^2$ are the mean value of $s$ and variance in $s$. The additional parameter $n$ represents the number of shocks giving rise to a distribution with a negative skew. More details, along with the explicit form for $\phi$ can be found in \citep{fsh23}.

By default, the finite shock model PDF without a superscript is assumed to describe the volume-weighted statistics of log density $s$. To obtain its mass-weighted counterpart, we employ \eqref{eq:PDF_weighting_MV1D}
\begin{equation}
    f^{(M)}_s (s; \mu, \sigma, n) = e^s \fsh (s; \mu, \sigma, n)
\end{equation}

The kinetic energy-weighted PDF of log density is derived in sec. \ref{sec:energy_density_PDF}.

\subsection{Generating function of the finite shock model}

For the purposes of calculating various expectation values within the finite shock model, we introduce the following parametric expectation value involving only (log) density
\begin{equation}
    E (u, k; \mu, \sigma, n) \equiv \left\langle s^k e^{u s} \right\rangle = \int \limits_{-\infty}^\infty s^k e^{u s} \fsh (s; \mu, \sigma, n)
    \label{eq:fshock_master_formula}
\end{equation}

Using the analytic properties of the characteristic function, we can easily calculate the expectation value for $k = 0$. Moreover, differentiation with respect to $u$ brings down one power of $s$, increasing $k$ by 1, which gives rise to a recurrent formula for $k \geq 1$,
\begin{align}
    E (u, 0; \mu, \sigma, n) &= e^{u \mu} \phi (- i u \sigma; n) \\
    E (u, k+1; \mu, \sigma, n) &= \frac{\D}{\D u} E (u, k; \mu, \sigma, n)
\end{align}

In order to extract useful quantities from the characteristic function, we introduce two normalized functions, $\Phi_k(x)$ and $F(\Delta)$, which normalize out the first and second arguments of $\phi(\omega;n)$ as follows
\begin{align}
    \label{eq:Phi0}
    \Phi_0 (x) &\equiv \frac{1}{n} \log \phi (- i \sqrt{n} x; n) \\
    \label{eq:FDelta}
    F (\Delta) &= \frac{1}{\sigma^2} \log \phi (- i \sigma; \sigma^2 / \Delta^2).
\end{align}

If $\mu, \sigma, n$ are parameters of the volume-based distribution of log density, the conservation of total mass, $\langle e^s \rangle = 1$, following equations \eqref{eq:fshock_master_formula} and \eqref{eq:Phi0}, constraints $\mu$ as follows
\begin{equation}
     \mu = - \log \phi (- i \sigma; n) = - n \, \Phi_0 (\sigma / \sqrt{n})
     \label{eq:muV}
\end{equation}
which, as expected, reduces to $- \sigma^2 / 2$ when $n \to \infty$.

The number of shocks, $n$, for given values $\mu, \sigma$ can be estimated from equation \eqref{eq:muV} and by inverting equation \eqref{eq:FDelta}
\begin{equation}
    n = \frac{\sigma^2}{\Delta \left(- \mu / \sigma^2 \right)}
    \label{eq:nV_from_mu_sigma}
\end{equation}
where $\Delta \equiv F^{-1}$ denotes the solution to equation \eqref{eq:FDelta}.

$\Phi_k(x)$ for $k > 0$ are calculated as the derivative of $\Phi_0$, and their explicit form for $k = 1,2$ is
\begin{align}
    \label{eq:Phi1}
    \Phi_1 (x) &\equiv \Phi^\prime (x) = - \frac{i}{\sqrt{n}} \frac{\phi^\prime (- i \sqrt{n} x; n)}{\phi (- i \sqrt{n} x; n)} \\
    \label{eq:Phi2}
    \Phi_2 (x) &\equiv \Phi^{\prime \prime} (x) = - \frac{\phi^{\prime \prime} (- i \sqrt{n} x; n)}{\phi (- i \sqrt{n} x; n)} + \left( \frac{\phi^\prime (- i \sqrt{n} x; n)}{\phi (- i \sqrt{n} x; n)} \right)^2
\end{align}

The mass-weighted counterpart of the average log density, ${\mu_M \equiv \langle s \rangle_M = \langle \rho s \rangle}$ can be calculated using the generating function $E$ with $u = 1, k = 1$, utilizing equation \eqref{eq:Phi1},
\begin{equation}
    \mu_M = \mu + \sqrt{n} \, \sigma \, \Phi_1 (\sigma / \sqrt{n})
    \label{eq:muM}
\end{equation}
reducing to $+ \sigma^2 / 2$ when $n \to \infty$.

Finally, it is possible to express the variance in $s$ weighted by mass, $\sigma_M^2 = \left\langle \rho s^2 \right\rangle - \left\langle \rho s \right\rangle^2$, using equation \eqref{eq:Phi2} as follows
\begin{equation}
    \sigma_M^2 = \sigma^2 \, \Phi_2 (\sigma / \sqrt{n})
    \label{eq:sigmaM}
\end{equation}
which reduces to $\sigma_M = \sigma$ in the lognormal limit.

\subsection{Energy-weighted density PDF}\label{sec:energy_density_PDF}

\begin{figure} \begin{center}
    \includegraphics[width=0.49\textwidth]{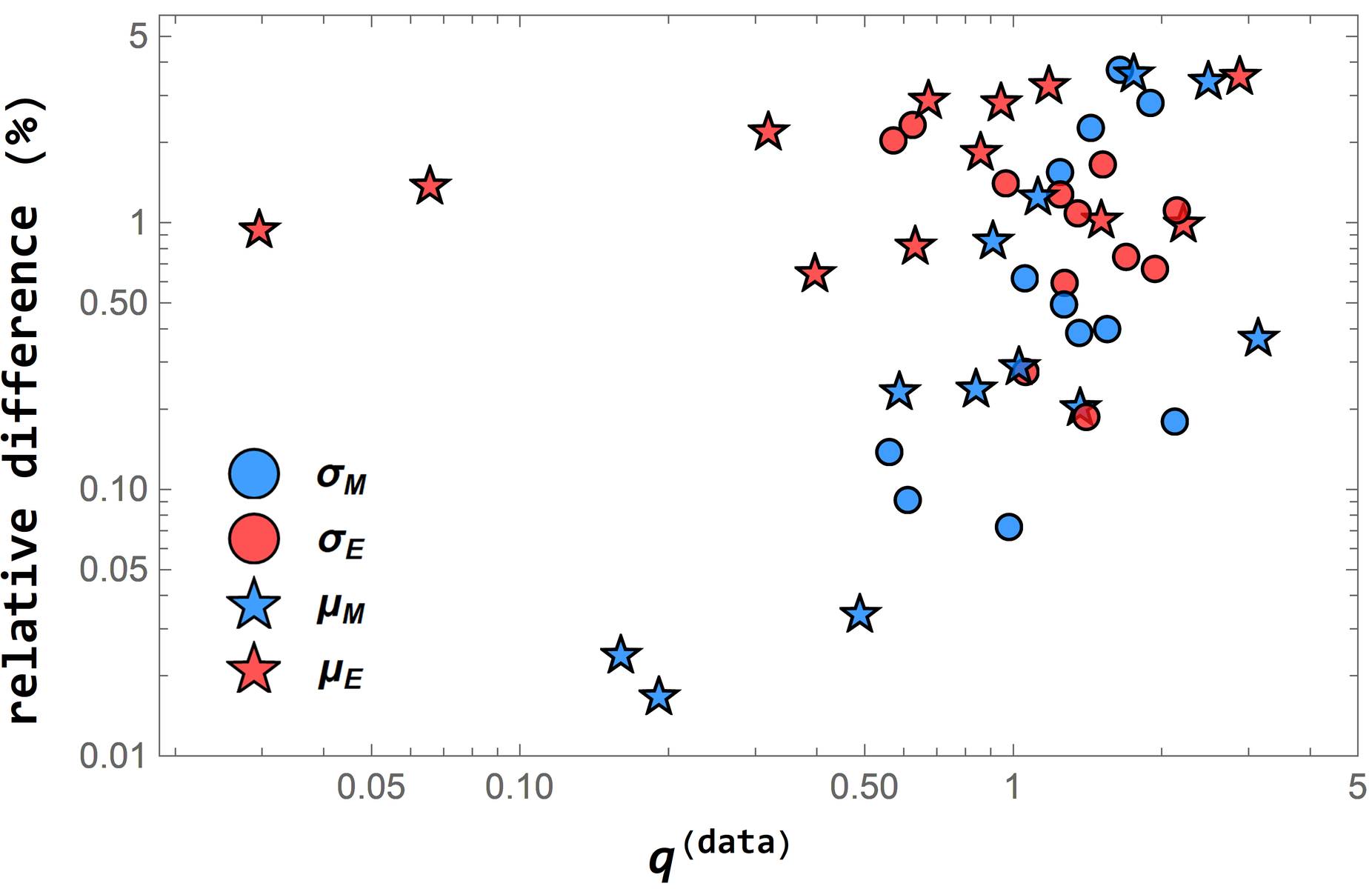}
\caption[ ]{Mass- and energy-weighted mean $\mu$ and variance $\sigma$ of log density. The horizontal axis represents the values as measured from each simulation. Vertical axis represents the relative error between the measured and theoretically predicted $\sigma$ (circles) and $\mu$ (stars) weighted by mass (blue) or energy (red). Since the typical scale within Gaussian-like distributions is set by the width $\sigma$, both errors in $\mu$ and $\sigma$ are considered relative to $\sigma_{M,E}$. The theoretical values are calculated using the values of $\mathfrak X, \mu, \sigma$ taken from the simulations. The number of shocks $n$ is obtained via equation \eqref{eq:nV_from_mu_sigma} and subsequently $\mu_M$ and $\sigma_M$ are calculated from equations (\ref{eq:muM}, \ref{eq:sigmaM}). Values $\mu_E, \sigma_E$ are approximated via equations (\ref{eq:sigmaE_approx}, \ref{eq:muE_approx}).}
\label{fig:estimatorsME} \end{center} \end{figure}

\begin{figure*} \begin{center}
    \includegraphics[width=0.99\textwidth]{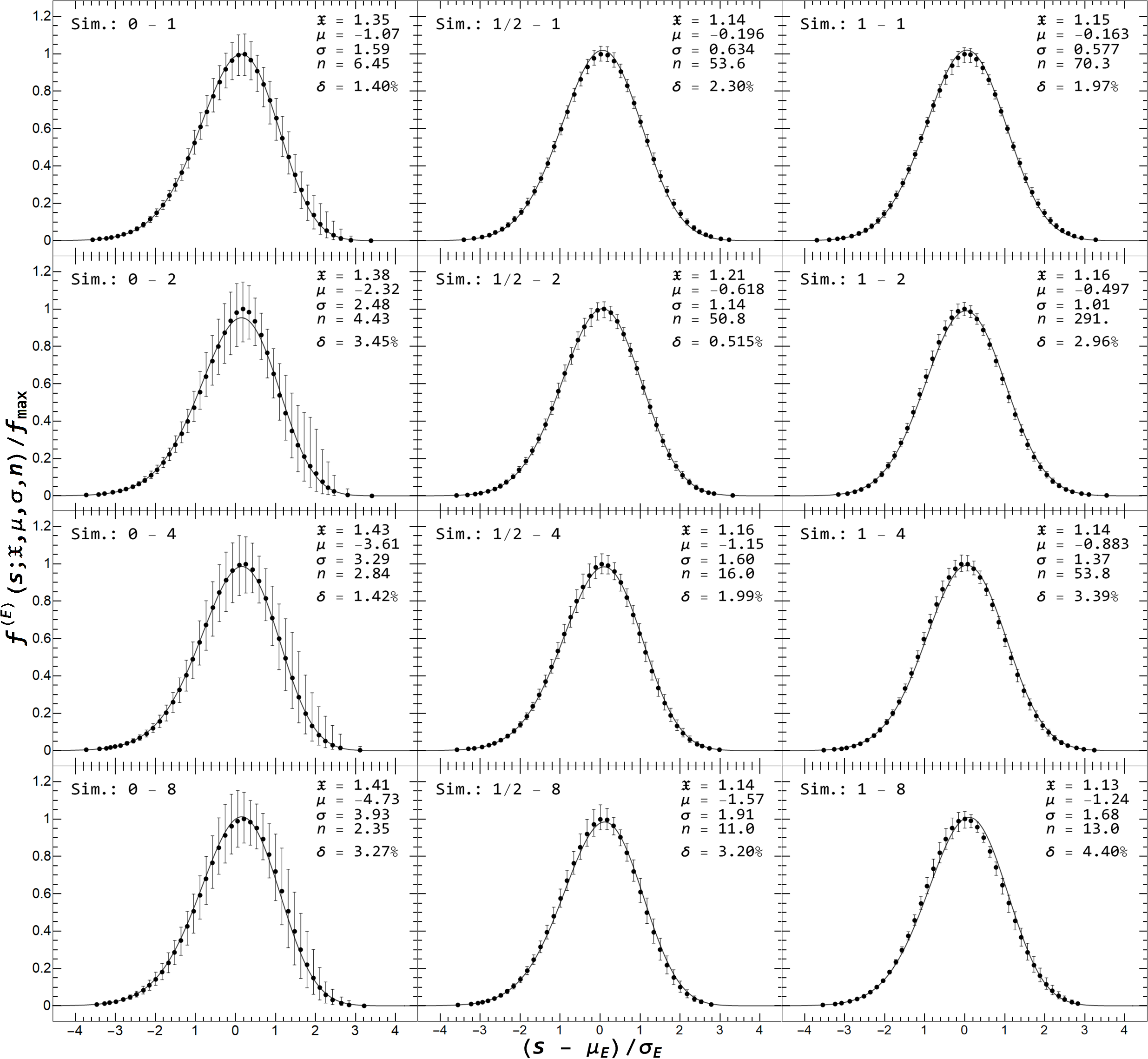}
\caption[ ]{Plots of the PDF (vertical axis) of log density (horizontal axis) weighted by kinetic energy. Each simulation is labeled with $\xi - M$ in the top left corner. Both axes are shifted and rescaled to emphasize the shape of the histograms. Data points (dots) along with the error bars (vertical lines) are represented in black. Measured parameters $\mathfrak X = M^2 / M_M^2, \mu, \sigma$ and $n$ using equation \eqref{eq:nV_from_mu_sigma} (listed in the top right corner of each plot) give rise to the solid black line.}
\label{fig:logrhoEfits} \end{center} \end{figure*}

For the construction of the joint PDF of density and speed as outlined in sec. \ref{section:joint}, the kinetic energy-weighted histogram of density must be known. We already explored the mass-weighted PDF, $f^{(M)}_s (s; \mu, \sigma, n) = e^s \fsh (s; \mu, \sigma, n)$, and its statistics in the previous paragraph. However, equation \eqref{eq:PDF_weighting_EM2D} indicates, that the conversion from the mass-weighted to the energy-weighted instance of the density PDF would require marginalization of the full joint PDF weighted by a factor of $v^2$. Since the full PDF is not known, this approach is not feasible. To sidestep this problem, we propose an explicit form for the energy-weighted PDF based on the finite shock model. First, we notice an approximate relation between the mass- and energy-weighted standard deviations of $\log \rho$ are approximately equal,
\begin{equation}
    \sigma_E \approx \sigma_M
    \label{eq:sigmaE_approx}
\end{equation}
to a high degree of accuracy. The highest relative difference between the two is observed to be less than $3\%$ in the compressive simulation with Mach number 2 (see Figure \ref{fig:estimatorsME}). This remarkable match allows for the following educated guess; since the \emph{width} of the log density PDF does not change between the mass- and energy-weighted instances, we assume, that the two share the same general shape. The only freedom left after this assumption has been made is an arbitrary argument shift, that can be expressed as
\begin{equation}
    f^{(E)}_s (s) = f^{(M)}_s (s + \delta s) = e^{s + \delta s} \fsh (s + \delta s; \mu, \sigma, n).
\end{equation}

As a consequence, the difference between the mean of $s$ weighted by energy and mass is $\delta s$; $\mu_M - \mu_E = \delta s$. To determine $\delta s$ we look at the energy-weighted mean of $1/\rho$,
\begin{equation}
    \langle e^{-s} \rangle_E = \frac{\langle v^2 \rangle}{\langle e^s v^2 \rangle} = \mathfrak X
\end{equation}
where $\mathfrak X = \langle v^2 \rangle / \langle \rho v^2 \rangle$ was introduced in equation \ref{eq:defX}.

Going back to our proposed shape for $f^{(E)}_s$, we use this newly found mean value to determine $\delta s$
\begin{equation}
    \mathfrak X = \langle e^{-s} \rangle_E = \int \limits_{-\infty}^\infty \D s \, e^{-s} f^{(E)}_s (s) = e^{\delta s} \; \implies \; \delta s = \log \mathfrak X
\end{equation}
which translates to the following shift in $\mu_E$
\begin{equation}
    \mu_E = \mu_M - \log \mathfrak X
    \label{eq:muE_approx}
\end{equation}

Given the shift, the energy-weighted PDF can be written using the finite shock model as
\begin{equation}
    f^{(E)}_s (s; \mathfrak X, \mu, \sigma, n) = \mathfrak X \, e^s \fsh (s; \mu - \log \mathfrak X, \sigma, n)
    \label{eq:fE}
\end{equation}

Figure \ref{fig:estimatorsME} shows the relative error between the estimators for $\sigma_{M, E}$ and the values measured from the simulations as filled circles. The calculated value for $\sigma_M$ was obtained from $\mu, \sigma, n$ using equation \eqref{eq:sigmaM}, where $n$ is given by equation \eqref{eq:nV_from_mu_sigma}. Subsequently, $\sigma_E$ is assumed to be equal to $\sigma_M$ per equation \eqref{eq:sigmaE_approx}. Figure \ref{fig:estimatorsME} also shows the error between the estimated and measured means $\mu_{M, E}$ (filled stars) obtained from equations (\ref{eq:muM}, \ref{eq:muE_approx}). These errors are taken relative to their respective $\sigma_{M,E}$, $\left| \mu_{M, E}^{(\text{data})} - \mu_{M, E}^{(\text{est.})} \right| / \sigma_{M, E}^{(\text{data})}$. This reduction was chosen due to the overall scale of a Gaussian-like distribution being set by its respective standard deviation $\sigma$; two Gaussian distributions with equal widths $\sigma$ only differ substantially from each other if their means $\mu$ disagree significantly on the scale given by $\sigma$. The difference between the estimated and measured mass- and energy-weighted values of mean and standard deviation of log density is below $5 \%$ for all simulations, demonstrating the accuracy and consistency of the approximations derived in this section.

Figure \ref{fig:logrhoEfits} shows the plots of $f^{(E)}_s (s; \mathfrak X, \mu, \sigma, n)$ compared to the histograms extracted from the simulations, by using the values of $\mathfrak X, \mu, \sigma$ directly measured from the histograms. These values are used to determine $n$ using equation \eqref{eq:nV_from_mu_sigma}. Subsequently, equation \eqref{eq:fE} with the determined parameters and the finite shock model for the volume-weighted basis is plotted alongside the data. The match between the model equipped by estimated parameters and the histograms is remarkable, considering the approximations made along the way.
\begin{figure*} \begin{center}
    \includegraphics[width=0.99\textwidth]{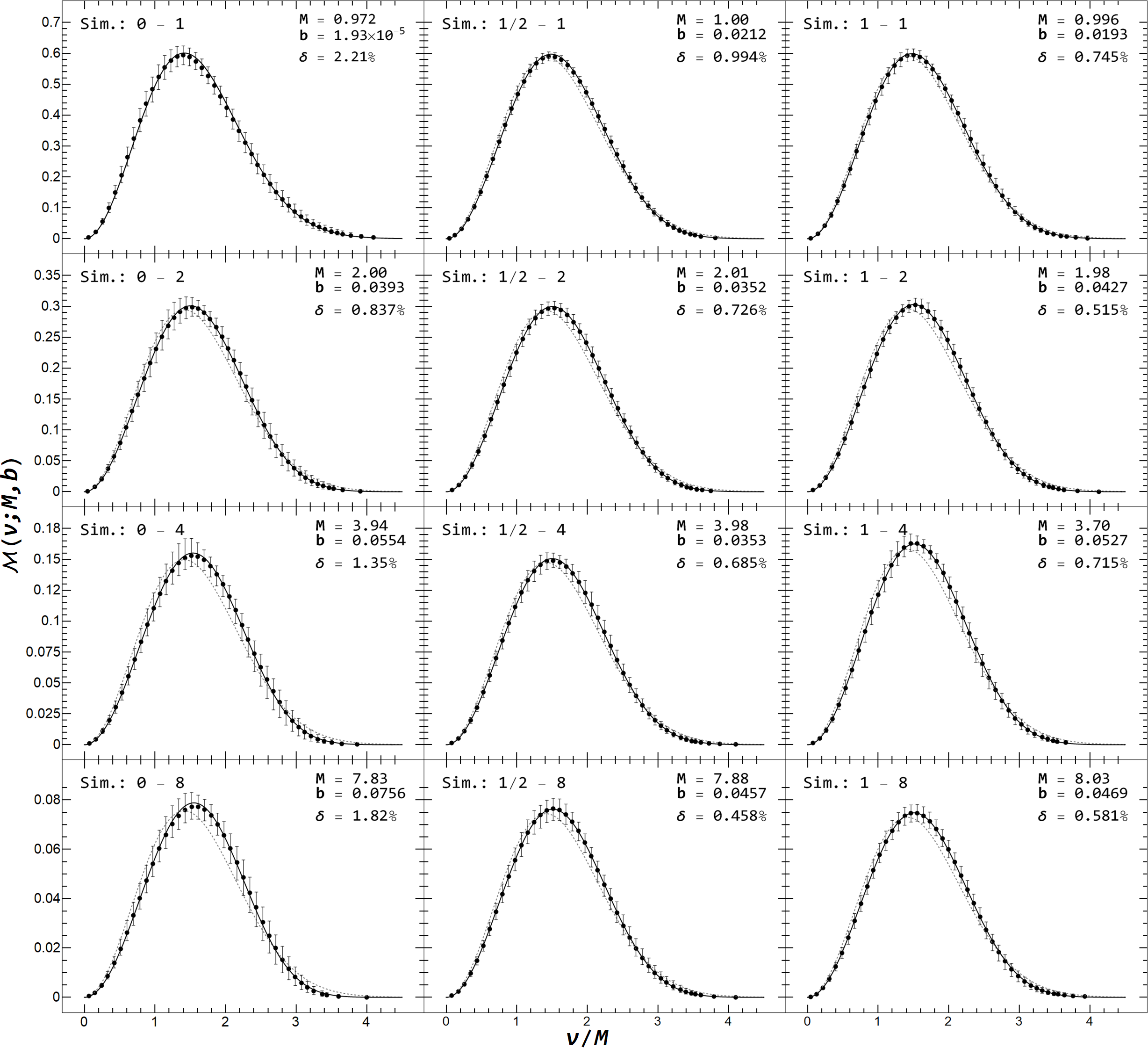}
\caption[ ]{Plots of the PDF (vertical axis) of speed (horizontal axis). Data points (dots) along with the error bars (vertical lines) are represented in black. The simple Maxwellian curve with the measured Mach number is shown as a dashed gray line, the Maxwellian with a quartic correction, obtained by measuring $v$ in addition to $M$, is represented by a solid black curve.}
\label{fig:absv} \end{center} \end{figure*}
\section{PDF of speed}\label{section:speed}

The velocity field within in isothermally turbulent medium can, due to the chaotic nature of turbulence, also be treated as a random variable with certain statistical properties. While the exact distribution depends on the driving, several assumptions can be made to derive a simple distribution for the magnitude of velocity.

Assuming independence of all components of velocity and isotropic driving, the argument similar to that of \citet{Maxwell} can be used to infer that the velocity is a Gaussian in all directions with variance equal in each component. Thus, the speed is drawn from the following Maxwellian distribution
\begin{equation}
    f_v (v; M) = \mathcal M (v; M) = \frac{4 \pi v^2}{(2 \pi c_s^2 M^2)^{3/2}} \exp \left( - \frac{v^2}{2 c_s^2 M^2} \right),
\end{equation}
where $M$ is the 1D r.m.s. Mach number.

%
In what follows we will set $c_s = 1$ for the sake of brevity.

Despite the vast majority of literature regarding the velocity fluctuations focuses on the two-point statistics and power spectra, several previous works address the deviations from the ideal Maxwellian shape of the PDF of speed in compressible and incompressible isothermal turbulence \citep{nonGaussian98, Gotoh02, Wilczek2011}. The slope of the distribution above the maximum can be observed to be steepened compared to the ideal Maxwellian, and can be seen from a direct comparison, in Figure \ref{fig:absv}. The three-dimensional geometry of the simulation necessarily implies that the prefactor $v^2$ is preserved under very general assumptions about the original distribution for the velocity, $f (\vec{v}) \to f (v) \sim v^2 + \cdots$. Thus, this steepening can only be reflected as a higher-order term, for example a quartic correction inside the exponential,
\begin{equation}
    f^{(V,M)}_v (v; M, b) = \Max (v; M, b) \propto v^2 \exp \left[ - \frac{v^2}{2 a^2} \left( 1 - b + \frac{b v^2}{a^2} \right) \right]
    \label{eq:quartic_fit_VM}
\end{equation}
where $a$ is a parameter carrying the units of speed, that is adjusted so that the root-mean square of $v$ matches the desired Mach number, $3 M^2 = \langle v^2 \rangle$. The parameter $b \in [0, 1]$ adjusts the amount of steepening; when $b = 0$, ideal Maxwellian shape is restored, whereas for $b = 1$, the tail behaves like $\sim v^2 e^{- v^4}$.


Note, that the functional form of equation \eqref{eq:quartic_fit_VM} can be used to describe both volume- and mass-weighted PDF of speed, with unique parameters of $M, b$ in each case. The kinetic energy-weighted histogram of speed can be determined using equation \eqref{eq:PDF_weighting_EM1D}.

The difference between the newly introduced correction and its Maxwellian counterpart when $b = 0$, apart from the shape of the PDF, manifests in the following ratio of the expectation values of powers of magnitude of speed
\begin{equation}
    \frac{\left\langle \left( \vec{v} \cdot \vec{v} \right)^\alpha \right\rangle}{\left\langle \left( \vec{v} \cdot \vec{v} \right)^\alpha \right\rangle_{(b=0)}} \equiv h_\alpha (b).
    \label{eq:stretching_factor}
\end{equation}
The function $h_\alpha$ only depends on the power, $\alpha$, and the tilt parameter, $b$.  While it doesn't have an analytic form, can be easily tabulated and inverted numerically.

Specifically, for the pure Maxwellian, the expected results are
\begin{equation}
    \left\langle \left( \vec{v} \cdot \vec{v} \right)^\alpha \right\rangle_{(b=0)} = \int \limits_0^\infty v^{2 \alpha} f_v (v; M) \, \D v = \frac{2^{\alpha + 1}}{\sqrt{\pi}} M^{2 \alpha} \Gamma (\alpha + 3/2)
    \label{eq:Maxwellian_expectations}
\end{equation}
which simplifies to $(2n+1)!! M^{2n}$ for integer $\alpha = n$, however, extra care should be taken for half-integer $\alpha$, as the double-factorial formula does not match the form in equation \eqref{eq:Maxwellian_expectations}. Lower values of $\alpha$ are most numerically reliable, for example, for $\alpha = 1/2$, we can relate the ensemble average of $\langle v \rangle$ to the sloping parameter $b$ as follows
\begin{equation}
    \sqrt{\frac{\pi}{8}} \frac{\left\langle v \right\rangle}{M} = \frac{\sqrt{\pi/8}}{M} \left\langle \sqrt{\vec{v} \cdot \vec{v}} \right\rangle = h_{1/2} (b) \; \to \; b = h_{1/2}^{-1} \left( \sqrt{\frac{\pi}{8}} \frac{\left\langle v \right\rangle}{M} \right)
    \label{eq:Mv_b_relation}
\end{equation}

This equation can be used to estimate the value of the parameter $b$ for a given set of measured ensemble averages $v$ and the Mach number $M$. Table \ref{tab:simpars} lists the simulation parameters along with the ensemble averages of $v$ and Mach number (both volume- and mass-weighted). Figure \ref{fig:absv} shows the perfect Maxwellian shape by obtaining the Mach number $M$ and the correction \eqref{eq:quartic_fit_VM} obtained by measuring the additional parameter $v \equiv \langle v \rangle$ for each simulation. While the Maxwellian form fails to fit the data for $v > M$ due to the prominent steepening of the slope of the distribution in this region, the quartic correction approximates the dataset much better.



In the line of the original argument for the Maxwellian distribution of speeds based on the rotational symmetry and independence of individual components of velocity, one might wonder which assumption (if not both) is violated. Arguments from the power spectrum of velocity \citep{nonGaussian98} and direct numerical simulations \citep{Wilczek2011} show, that the tails of the PDFs of the individual components of velocity are sub-Gaussian, which does not leave any indication of dependence or independence of the components. The full study of the velocity statistics is interesting, but outside the scope of this work.
\begin{figure*} \begin{center}
    \includegraphics[width=0.99\textwidth]{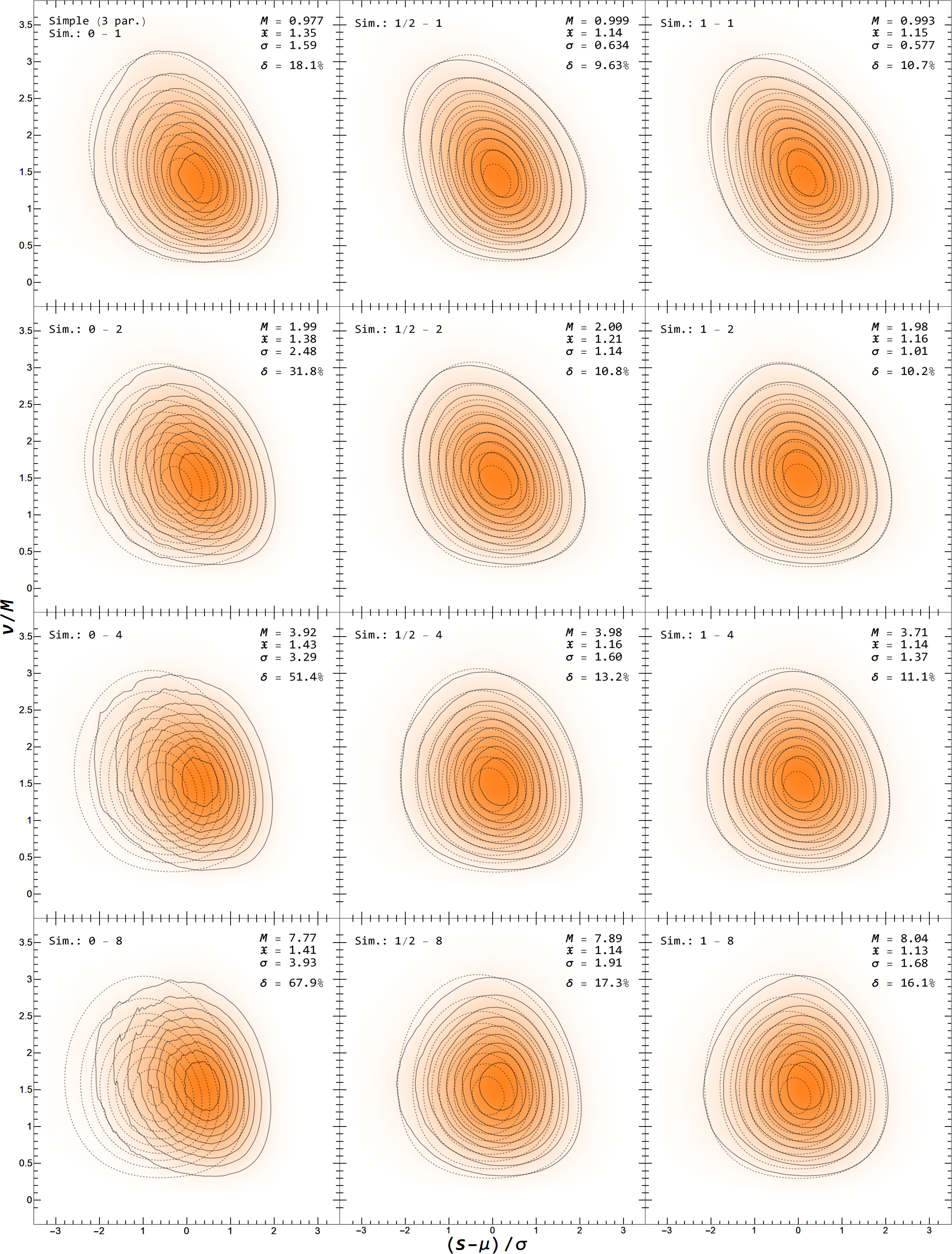}
\caption[ ]{Plot of the volume-weighted PDF of log density (horizontal axis) and speed (vertical axis), solid contours, compared to the minimal model with 3 parameters (listed on the plot) obtained as ensemble averages. The axes are shifted and rescaled for the sake of clarity.}
\label{fig:simple3_grid} \end{center} \end{figure*}
\begin{figure*} \begin{center}
    \includegraphics[width=0.99\textwidth]{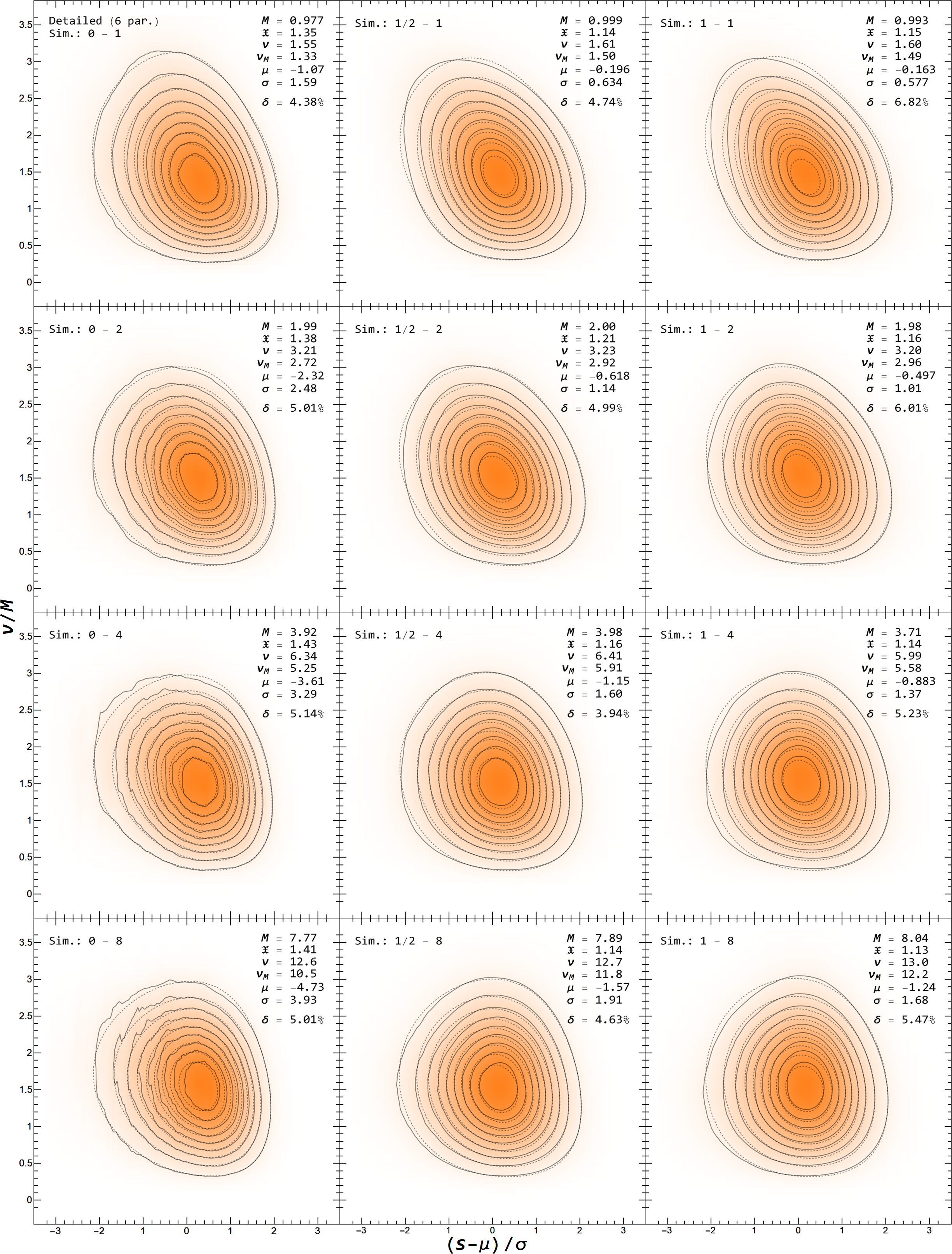}
\caption[ ]{Plot of the volume-weighted PDF of log density (horizontal axis) and speed (vertical axis), solid contours, compared to the detailed model featuring 6 parameters obtained as ensemble averages (listed on each plot). The axes are shifted and rescaled for the sake of clarity.}
\label{fig:elaborate6} \end{center} \end{figure*}
\section{Joint PDF of density and speed: general theory}\label{section:joint}

We now turn to the joint distribution of density and speed, $f_{(s,v)}(s,v)$. Having already described the statistics of each variable separately, the dependence between the two comes to question, as
\begin{equation}
	(s, v) \text{\; independent} \; \iff \; f_{(s,v)} (s,v) = f_s (s) f_v (v).
\end{equation}
If the random variables are truly independent, the joint PDF would be fully described by the product of its marginalized parts, ${f_{(s,v)} (s,v) = f_s (s) f_v (v)}$.  Conversely, if there is dependence between $s$ and $v$, $f_{(s,v)}$ is not the product of the marginalized distributions.  We will first show that this is in fact the case, then develop a model for the actual joint PDF.  Our correction will be developed in the next section.

To demonstrate dependence between $s$ and $v$, we exploit another, equivalent, definition of independence of random variables. For any two functions $h_1(s), h_2(v)$: ${\left\langle h_1 (s) \, h_2 (v) \right\rangle = \left\langle h_1 (s) \right\rangle \left\langle h_2 (v) \right\rangle}$ iff $s, v$ are independent random variables. That is, the average of the product is the product of the averages, iff $s$ and $v$ are independent.  Conversely, if we find a certain combination for which $\left\langle h_1 (s) h_2 (v) \right\rangle \neq \left\langle h_1 (s) \right\rangle \left\langle h_2 (v) \right\rangle$, the variables must be dependent.

One such choice is $h_1(s)=s$ and $h_2(v)=v^2$.   We will show that $\left\langle \rho v^2 \right\rangle \neq \left\langle \rho \right\rangle \left\langle v^2 \right\rangle$.  We can interpret this as the mass-weighted r.m.s. Mach number, also related to the mean kinetic energy density $\varepsilon$,
\begin{equation}
    \varepsilon = E / V = \left\langle \frac{1}{2} \rho v^2 \right\rangle = \frac{3}{2} \rho_0 M_M^2
\end{equation}
where $E$ is the total kinetic energy, $E = \varepsilon V$.  We parameterize the correlation using $\mathfrak X=\langle v^2\rangle/\langle \rho v^2\rangle$ and show that it is different from one, demonstrating dependence.

Table \ref{tab:simpars} features all parameters measured from the simulations. As seen from the values of $\mathfrak X$, the values of $\langle v^2 \rangle$ and $\langle \rho v^2 \rangle$ differ by at least 10\% in all simulations which indicates, that density and speed are correlated and therefore, to some extent, dependent quantities.

The non-zero correlation between density and speed complicates the joint statistics, since the joint PDF cannot be written as a product of the 1D marginalized PDFs. However, motivated by the fact, that the product of 1D marginalized PDFs is relatively close to the joint PDF, in the following section \ref{subsec_correction} we propose a simple correction term added to the product of marginalized distributions, allowing for a simple, consistent, description of the joint statistics.

\subsection{Correction term to the joint PDF}\label{subsec_correction}

The relative proximity between the true joint PDF and the product of its marginalized subparts leads us to believe, that a simple, small correction to the latter can be used to model the dependence between $s$ and $v$,
\begin{equation}
	f_{(s,v)} (s,v) = f_s (s) f_v (v) + g (s, v).
\end{equation}

Given full freedom in $g$, this approach can perfectly describe the joint PDF. However, the full knowledge of such correction is akin to knowing the joint PDF itself. Instead, we resort to a reasonable approximation; let's assume, that the function $g$ can be also written as a product of two single-variable functions,
\begin{equation}
	g (s, v) = g_s (s) g_v (v).
        \label{eq:g_separation}
\end{equation}

The main task is to determine the single variable functions $g_{s, v}$ using various methods of weighting outlined in \ref{sec:weighting}. Note, that since integrating out one of the variables must yield the marginalized PDF of the other variable, the integral over each single-variable $g_{s,v}$ must be equal to zero. Therefore, to reveal the correction term in each variable, we need a way to break this symmetry by introducing a factor involving one of the variables. This can be done using the paradigm of weighted histograms, as weighting by different positive quantities naturally imposes factors involving density and speed.

To proceed, we consider the mass-weighted joint PDF of $s$ and $v$ as the basis for our calculations,
\begin{align}
    f_{(s,v)}^{(M)} (s, v) = f_s^{(M)} (s) f_v^{(M)} (v) + g_s^{(M)} (s) g_v^{(M)} (v), \label{equation:model}
\end{align}
and compare it to the volume- and kinetic energy-weighted joint PDFs, that can be related to the mass-weighted basis using equations (\ref{eq:PDF_weighting_MV2D}, \ref{eq:PDF_weighting_EM2D})
    \begin{align}
        f_{(s,v)}^{(V)} (s, v) &= e^{-s} \left[ f_s^{(M)} (s) f_v^{(M)} (v) + g_s^{(M)} (s) g_v^{(M)} (v) \right] \label{equation:fv} \\
        f_{(s,v)}^{(E)} (s, v) &= \frac{v^2}{3 M_M^2} \left[ f_s^{(M)} (s) f_v^{(M)} (v) + g_s^{(M)} (s) g_v^{(M)} (v) \right] \label{equation:fe}
    \end{align}

The factors introduced this way break the symmetry of the correction terms under integration over the involved variable. Firstly, by definition, integrating over the mass-weighted instances of the joint PDF yields the baseline mass-weighted marginalized distribution of the other variable
\begin{align}
    \int \limits_{-\infty}^\infty \D s \; f_{(s,v)}^{(M)} (s, v) &= f_v^{(M)} (v) \\
    \int \limits_0^\infty \D v \; f_{(s,v)}^{(M)} (s, v) &= f_s^{(M)} (s)
\end{align}

If we now use the fact, that $\left\langle e^{-s} \right\rangle_M = \langle 1 \rangle = 1$ and $\left\langle v^2 \right\rangle_M = \langle e^s v^2 \rangle = 3 M_M^2 = 2 \varepsilon$, we can explicitly integrate out $s$ in the volume-weighted case and $v$ in the energy-weighted instance to get
    \begin{align}
        \int \limits_{-\infty}^\infty \D s \; f_{(s,v)}^{(V)} (s, v) &\equiv f_v^{(V)} (v) = f_v^{(M)} (v) + A g_v^{(M)} (v) \label{equation:ms}\\
        \int \limits_0^\infty \D v f_{(s,v)}^{(E)} (s, v) &\equiv f_s^{(E)} (s) = f_s^{(M)} (s) + B g_s^{(M)} (s) \label{equation:mv}
    \end{align}
where $A, B$ are non-zero constants associated with the integrals of the mass-weighted $g$-functions of variable $s$ and $v$ with additional factors of $e^{-s}$ and $v^2$ in the density and speed terms, respectively. As we can see, the terms associated with different weighing break the symmetry of an otherwise identically vanishing integral. Solving equations \eqref{equation:ms} and \eqref{equation:mv} for the $g-$functions, we find:
\begin{align}
    g_s^{(M)} (s) &\sim f_s^{(E)} (s) - f_s^{(M)} (s)\\ g_v^{(M)} (v) &\sim f_v^{(V)} (v) - f_v^{(M)} (v).
\end{align}
The corrected joint PDF of $\log \rho$ and $v$ can be found by inserting these into equation \eqref{equation:model} to find
\begin{multline}
    f_{(s,v)}^{(M)} (s,v) = f_s^{(M)} (s) f_v^{(M)} (v) \, + \\
    + C \Big( f_s^{(E)} (s) - f_s^{(M)} (s) \Big) \Big( f_v^{(V)} (v) - f_v^{(M)} (v) \Big)
    \label{gfun_final_M}
\end{multline}
where $C$ is a constant accommodating the proportionality relation of the $g$-terms to the differences in the brackets.

This method is successful under two conditions; first, we had to assume, that the correction $g$ can be written as a product of two single-variable functions. Second, the single-variable functions must be well described by the finite shock model function and tilted Maxwellian, for some choice of the parameters, regardless of the method of weighing. It should be noted, that despite the derivation mainly focusing on the mass-weighted version of the histogram, this functional form can be converted to the volume-weighted instance of the joint PDF by multiplying by a factor $e^{-s}$. Since $f^{(M)}_s (s) = e^s f^{(V)}_s (s)$, we can write the volume-weighted joint PDF as follows
\begin{multline}
    f_{(s,v)}^{(V)} (s,v) \approx f_s^{(V)} (s) f_v^{(M)} (v) \, + \\
    + C \Big( e^{-s} f_s^{(E)} (s) - f_s^{(V)} (s) \Big) \Big( f_v^{(V)} (v) - f_v^{(M)} (v) \Big)
    \label{gfun_final_V}
\end{multline}

The expression for $C$,
\begin{equation}
    C = (\mathfrak X - 1)^{-1},
    \label{eq:relations_X}
\end{equation}
can be found by multiplying equation \eqref{gfun_final_M} by $v^2$, integrating over speed and demanding both sides to be equal to $3 M_M^2 f^{(E)}_s (s)$.

\section{Joint PDF: Specific Realizations}
\label{section:specific}

In what follows we suggest several choices of basis functions to build up the joint distribution; first, we use the simplest basis possible, consisting of Gaussian in $s$ and Maxwellian in $v$. We then utilize our updated marginalized pictures using the finite shock model and a tilted Maxwellian to obtain a much better description of the joint distribution.

\subsection{Minimal model}

In this section we describe the joint PDF using the simplest basis distributions; the normal distribution $\mathcal{N} (s; \mu, \sigma)$ with a mean $\mu$ and variance $\sigma^2$, and a simple Maxwellian $\mathcal{M} (v; M)$ where $M$ is the 1D r.m.s. Mach number. The minimum amount of parameters needed to describe the distribution is 3; $M, \mathfrak X, \sigma$. These three allow to directly describe the volume-weighted distribution of density, approximated by $\mathcal{N} (s; \mu, \sigma)$ where $\mu = -\sigma^2 / 2$, volume-weighted distribution of speed approximated by $\mathcal{M} (v; M)$ and also the mass-weighted distribution of speed using the Maxwellian with the parameter $M_M = M / \sqrt{\mathfrak X}$. The energy-weighted distribution of log density is approximated as $\exp (s + \log \mathfrak X) \mathcal{N} (s; \mu - \log \mathfrak X, \sigma)$. With these considerations in mind, the joint PDF can be then written as
\begin{multline}
    \begin{aligned}
        f^{(V)}_{(s,v)} &(s,v; M, \mathfrak X, \sigma) = \mathcal{N} (s; \mu, \sigma) \mathcal{M} (v, M_M) \, + \\
        &+ (\mathfrak X - 1)^{-1} \Big( \mathfrak X \, \mathcal{N} (s; \mu - \log \mathfrak X, \sigma) - \mathcal{N} (s; \mu, \sigma) \Big) \, \times \\
        &\times \, \Big( \mathcal{M} (v; M) - \mathcal{M} (v; M_M) \Big)
    \end{aligned}
\end{multline}

While this model does not aspire to fit the true shape of the 2D histogram, it fully preserves the measured parameters and expected relations between them.

Figure \ref{fig:simple3_grid} shows the joint PDF of $s$ (horizontal axis) and $v$ (vertical axis). Histograms obtained from the simulated data are displayed via solid contours and color denoting the fraction of probability, our minimal model of the joint PDF is overlaid as dashed contours. Since the minimal model only uses three parameters directly measured from the data, it cannot, in its simplicity, fully capture the joint PDF. The most jarring difference occurs in the compressively driven simulations with high r.m.s. Mach number, manifesting in a large shift of the maximum. This is due to a crude approximation $\mu = - \sigma^2 / 2$. In reality, $\mu$ is far away from this value, moreover, the true maximum of the density PDF is further shifted to the right due to the very low number of shocks inferred from these datasets.

While the maximum of the proposed simple model is shifted with respect to the true maximum of the distribution due to the approximations we used, the general shape matches that of the measured histograms.
\subsection{Detailed basis}

The final, most complicated form of our model of the joined distribution, we replace each function with its more detailed counterpart; the finite shock model function $\fsh (s; \mu, \sigma, n)$ instead of a simple Gaussian and the tilted Maxwellian for speed $\Max (v; M, b)$ in place of the ideal Maxwellian. This way, we need to provide 6 parameters to fully describe the joint distribution; $M, \mathfrak X, u, u_M, \mu, \sigma$, where $u, u_M$ are two new measured quantities equal to $u=\langle \sqrt{\vec{v} \cdot \vec{v}} \rangle$ and $u_M=\langle \sqrt{\vec{v} \cdot \vec{v}} \rangle_M = \langle \rho \sqrt{\vec{v} \cdot \vec{v}} \rangle$, which define $b$ and $b_M$ via equation \eqref{eq:Mv_b_relation}. Parameter $n$ is inferred from $\mu, \sigma$ using equation \eqref{eq:nV_from_mu_sigma}.

The function can then be written as
\begin{multline}
        f^{(V)}_{(s, v)} (s, v; M, \mathfrak X, u, u_M, \mu, \sigma) = \fsh (s; \mu, \sigma, n) \Max (v; M_M, b_M) \, + \\
        + (\mathfrak X - 1)^{-1} \Big( \mathfrak X \, \fsh (s; \mu - \log \mathfrak X, \sigma, n) - \fsh (s; \mu, \sigma, n) \Big) \, \times \\
        \times \Big( \Max (v; M, b) - \Max (v; M_M, b_M) \Big)
\end{multline}


Figure \ref{fig:elaborate6} shows the comparison between the model with detailed basis to the histograms extracted from the datasets. Notice the remarkable match between the two without any additional fitting. Even the noisiest dataset, the compressible Mach 8 simulation, is described very closely by our model in the regions with low noise and extrapolates naturally into the region with larger density and higher noise.

\section{Moments of the joint distribution}
\label{section:moments}

\begin{figure} \begin{center}
    \includegraphics[width=0.49\textwidth]{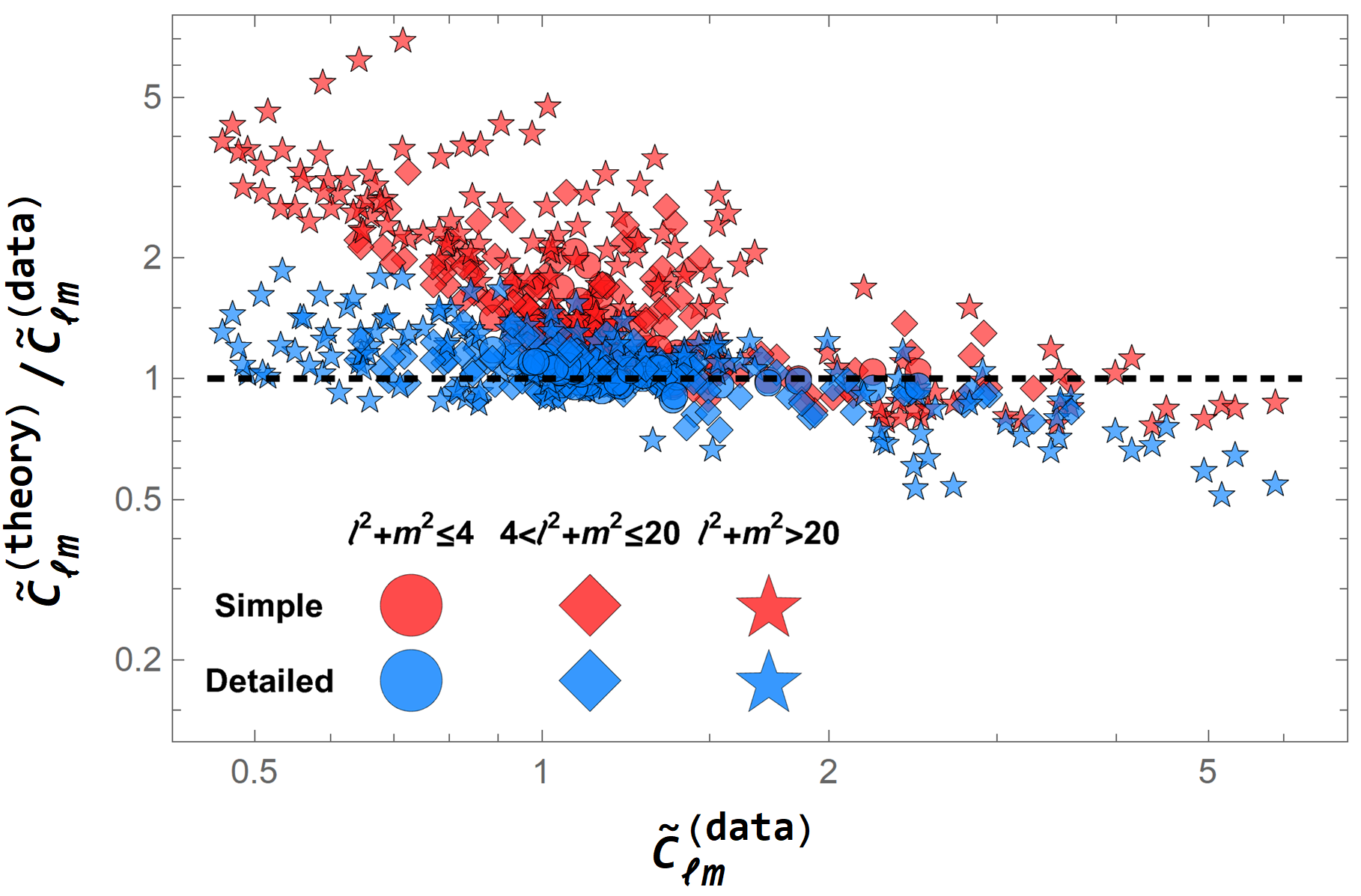}
\caption[ ]{Right panel: various correlators (horizontal axis) and their theoretical predictors reduced by its corresponding measured counterpart (vertical axis). Note, that all correlators are reduced by their theoretical value in case of uncorrelated density and speed, $\langle s^\ell \rangle \langle v^{2 m} \rangle$, for the sake of readability. The plot points are differentiated by color (model used to predict the correlator value) and shape (ranging from filled circles for the lowest values of $\ell, m$, diamonds for intermediate values of $\ell, m$ and triangles denoting the highest values of $\ell, m$).}
\label{fig:correlators} \end{center} \end{figure}

To corroborate our model of joint distribution, we compare various moments, $C_{\ell, m} = \langle s^\ell v^{2m} \rangle$ between our model and the data. The moments implied from our model can be expressed via the measured quantities as follows
\begin{equation}
    \begin{aligned}
        &C_{\ell,m} = (2m+1)!! M_M^{2m}\bigg[ E (0, \ell; \mu, \sigma, n) h_m (b_M) \, + \\
        & (\mathfrak{X} - 1)^{-1} \Big( \mathfrak X \, E (0, \ell; \mu - \log \mathfrak X, \sigma, n) - E (0, \ell; \mu, \sigma, n) \Big) \, \times \\
        & \Big( \mathfrak{X}^m h_{m} (b) - h_m (b_M) \Big) \bigg].
    \label{eq:correlators}
    \end{aligned}
\end{equation}

In case of the simple model using parameters $M, \mathfrak X, \sigma$, the correlators can be obtained from the same formula by taking ${n \to \infty}$, ${\mu = -\sigma^2 / 2}$ and $b = b_M = 0$. 

Figure \ref{fig:correlators} shows the ratio of the calculated vs. simulated moments of the joint distribution, $C_{\ell,m}$ for integers $1 \leq \ell, m \leq 5$. For the sake of clarity, all moments are normalized by their uncorrelated value assuming lognormal density and Maxwellian speed, ${\tilde C_{\ell, m} = C_{\ell, m} / ( \langle s^\ell \rangle \langle v^m \rangle )}$. Moments generated using the simple model are depicted by red points, those of the detailed model by blue points.  The shape of the points represents the size of $\ell^2+m^2$; the lowest powers are denoted by circles, intermediate powers by diamonds and the highest combinations of powers by stars.  It can be seen that for the most combinations of exponents, the detailed model matches the simulated moments substantially better than the simple model.

\subsection{Correlation coefficient}

The Pearson correlation coefficient $\text{corr} (s,v)$ is a special case of a normalized moment of the joint distribution and can be expressed using our model. The term $\langle s v \rangle$ needed to calculate $\text{corr} (s,v)$ can be obtained from equation \eqref{eq:correlators} by setting $\ell = 1, m = 1/2$,
\begin{equation}
    \text{corr} (s,v) = \frac{\langle s v \rangle - \langle s \rangle \langle v \rangle}{\sigma_s \sigma_v} = - \frac{u - u_M}{\sigma \sqrt{3 M^2 - u^2}} \frac{\mathfrak X \log \mathfrak X}{\mathfrak X - 1}
\end{equation}

\begin{figure} \begin{center}
\includegraphics[width=0.49\textwidth]{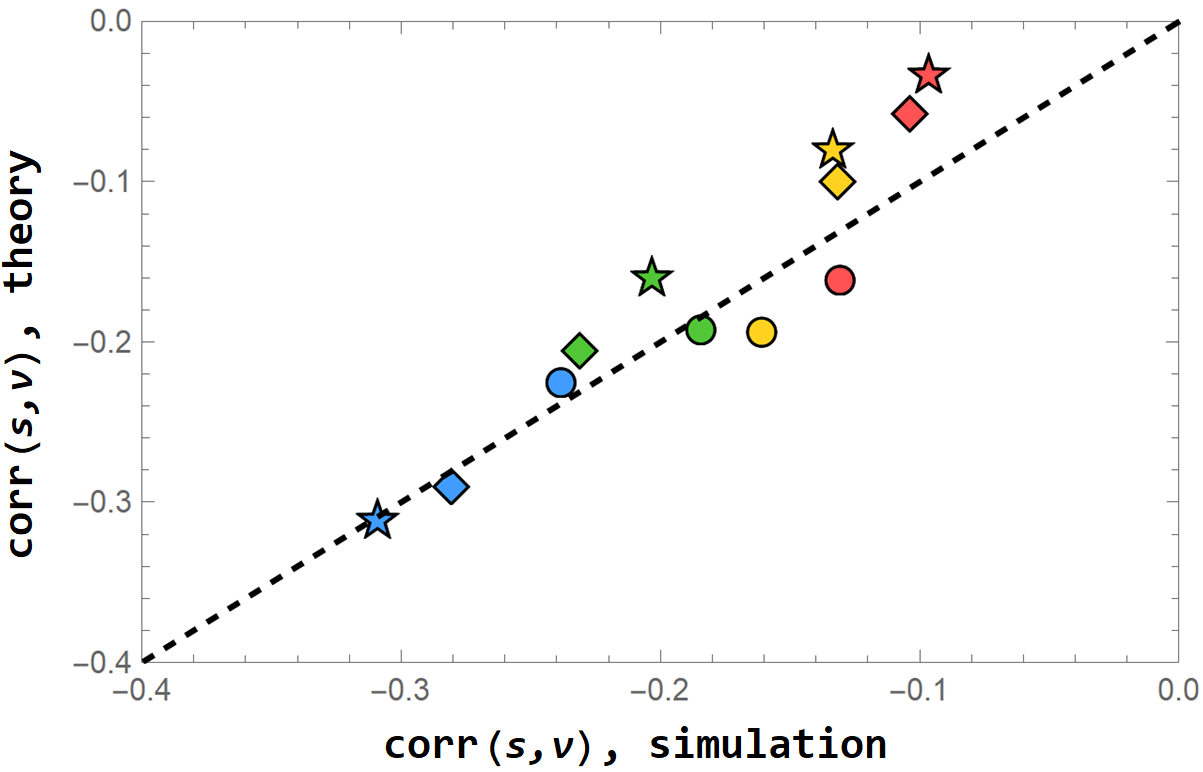}
\caption[ ]{Pearson correlation coefficient as measured from the simulations (horizontal axis) vs. the theoretical prediction (vertical axis). Different forcing modes are distinguished by the shape: disks for compressive, diamonds for mixed and stars for solenoidal forcing. The Mach number is distinguished by color: $M = 1$ (blue), 2 (green), 4 (yellow), 8 (red). The solid black line indicates $y = x$.}
\label{fig:pearson} \end{center} \end{figure}

This expression is compared to the measured correlation coefficients in Figure \ref{fig:pearson}. The largest correlation coefficients, occurring in the datasets with the lowest Mach numbers, match the measurement more accurately, whereas with increasing Mach number and decreasing correlation, the estimate of the correlation deviates somewhat from the measured value.

\section{Conclusions}\label{section:conclusions}

In the present work we developed a new model of the joint distribution of log density $s$ and speed $v$ by introducing a correction term to the product of marginalized 1D PDFs of the individual variables. By marginalizing over differently weighted instances of the proposed 2-dimensional PDF we were able to describe the correction term using a simple set of 1D distributions of each variable weighted by volume, mass or kinetic energy. We proposed 3 different shapes of the overall distribution, depending on the complexity of the basis functions; ranging from the simplest Gaussian in $s$ and Maxwellian in $v$ to the most detailed basis comprised of the finite shock model in $s$ and tilted Maxwellian (with a quartic correction) in $v$. Along the way we found out, that the kinetic-energy weighted histogram of log density has the same overall shape as its mass-weighted counterpart, and is shifted by $\delta s = \log \mathfrak X = \log \left( \langle v^2 \rangle / \langle \rho v^2 \rangle \right)$ to the left. The overall match between the shapes is closely related to the fact, that $\sigma_M = \sigma_E$, i.e. the mass- and kinetic energy-weighted variances of log density are equal to each other. The shift between the PDFs can be interpreted as the difference between the mass- and kinetic energy-weighted means of log density, $\mu_M - \mu_E = \log \mathfrak X$.

\begin{figure*} \begin{center}
\includegraphics[width=0.99\textwidth]{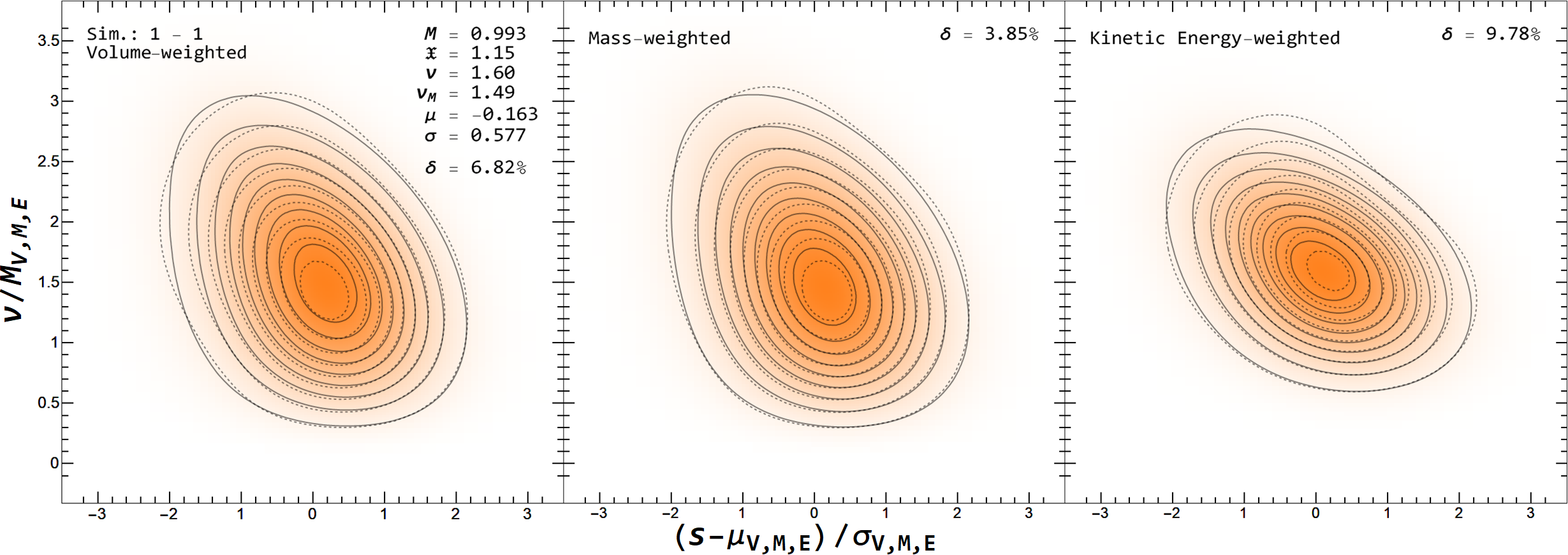}
\caption[ ]{The comparison between the histograms extracted from the simulated data with solenoidal forcing and $M = 1$ and our theoretical prediction, weighted by different quantities; volume (left), mass (center) and kinetic energy (right). Both axes are shifted and rescaled by their corresponding mean and standard deviation.}
\label{fig:VME} \end{center} \end{figure*}

Our model was confronted with simulated data from Enzo with compressive, mixed and solenoidal driving, each at 4 different 1D sonic Mach numbers $M = 1, 2, 4, 8$. The parameters of the model are directly measured from each simulation, with no additional fitting needed. The model using the detailed basis functions matches the simulated histograms to a high degree of precision even when density and speed are correlated to a considerable degree. The match between each model and histograms is measured by the $L_1$ norm, and for the detailed basis, the overall difference is at most $4.5\%$ in the worst case scenario. It should be noted, that feeding the model parameters taken from an ensemble leads to a reasonable match even upon re-weighing by mass or energy, e.g. see Figure \ref{fig:VME}. This is opposed to fitting one of the instances (for example the volume-weighted histogram) by varying the parameters of the model, however, that makes the match between a differently weighted histogram and its measured counterpart suboptimal.

In addition to matching histograms we computed a set of 25 correlation coefficients for each model, $\langle s^\ell v^{2 m} \rangle \; (1 \leq \ell, m \leq 5)$ that are compared to the coefficients measured directly from each simulation. Unsurprisingly, the model utilizing the detailed basis functions provides the closest match between the estimated values of the coefficients and their measured counterparts, with the factor of 2 at most, occurring in the case of the highest powers in $\ell, m$.

In this work we focused on the supersonic turbulent flows, in which the density and speed become less correlated with increasing Mach number, regardless of the forcing mode. At the same time, the number of shocks, inferred from the statistics of density alone, decreases with Mach number, resulting in a more tilted distribution. Both of these effects can be explained in the same framework of shocks and rarefaction waves. The shock waves propagating through a supersonic, turbulent medium exhibit, on average, higher density with increasing Mach number. However, due to overall mass conservation, the volume available for such shock to occupy is smaller, resulting in a limited longitudinal size of the shock wave. On the other hand, rarefaction waves, following behind the shocks, tend to reset the density towards the mean. Since the shock waves are faster and smaller in more turbulent gas, the number of shocks experienced by the gas before it resets to ambient density is smaller. This is paralleled by the weakening correlation between the density and speed.

Overall, our model suggests, that the correlations between density and speed are an integral part of the complete picture of the statistics of a turbulent, supersonic, isothermal flow. Moreover, with the knowledge of the full joint PDF of density and speed, further insight into the statistics of turbulence can be attained, such as exploring the statistics of thermal and kinetic energy.

\section*{Data Availability}
Simulation data present here is available on request (br18b@fsu.edu).

\section*{Acknowledgements}
\addcontentsline{toc}{section}{Acknowledgements}

Support for this work was provided in part by the National Science Foundation
under Grant AAG-1616026 and AAG-2009870.  Simulations were performed on \emph{Stampede2}, part of the Extreme Science and Engineering Discovery Environment \citep[XSEDE;][]{Towns14}, which is supported by National Science Foundation grant number
ACI-1548562, under XSEDE allocation TG-AST140008.

\bibliographystyle{mnras}
\bibliography{main.bib}



\bsp	
\label{lastpage}
\end{document}